\documentstyle[preprint,epsfig,aps,amssymb]{revtex}
\begin{document}
\draft
\tightenlines

\def\br{{\bf r}}
\def\bk{{\bf k}}
\def\bu{{\bf u}}
\def\bw{{\bf w}}
\def\brt{\br,t}
\def\bbrt{(\brt)}
\def\bprt{\bp,\brt}
\def\cphio{\Phi_0}
\def\beq{\begin{equation}}
\def\eeq{\end{equation}}
\def\bea{\begin{eqnarray}}
\def\eea{\end{eqnarray}}
\def\bna{\bbox{\nabla}}
\def\bp{{\bf p}}
\def\bv{{\bf v}}
\def\tn{\tilde n}
\def\tp{\tilde p}
\def\be{\bbox{\eta}}

\title{Dynamical Simulations of Trapped Bose Gases at Finite Temperatures}
\author{B. Jackson and E. Zaremba}
\address{Department of Physics, Queen's University, Kingston, Ontario 
 K7L 3N6, Canada.}
\date{\today}
\maketitle
\begin{abstract}

In this paper, we develop a numerical procedure for investigating the
dynamics of trapped Bose gases based on the ZGN theory. The dynamical
equations used consist of a generalized Gross-Pitaevskii equation for 
the condensate order parameter and a semiclassical kinetic equation 
for the thermal cloud. The former is solved using a fast Fourier
transform split-operator technique while the Boltzmann equation is
treated by means of $N$-body simulations. The two components are
coupled by mean fields as well as collisional processes that transfer
atoms between the two. This scheme has been applied to a model
equilibration problem, dipole oscillations in isotropic traps and
scissors modes in anisotropic traps. In the case of the latter, the
frequencies and damping rates of the condensate mode have been extracted
from the simulations for a wide range of temperatures. Good agreement
with recent experiments has been found.

\end{abstract}
\pacs{PACS numbers: 03.75.Fi, 05.30.Jp, 67.40.Db}

\section{Introduction}

The achievement in 1995 of Bose-Einstein condensation (BEC) in trapped atomic
gases \cite{anderson95,bradley95,davis95} 
initiated a period of intense experimental and theoretical 
research which continues at an ever-increasing pace. Due to the relative
diluteness of the atomic vapours, these systems present an excellent 
opportunity to investigate the effects of Bose degeneracy in novel situations.
For example, a rich array of dynamical behaviour can occur as a result
of the simultaneous occurence of a condensate order parameter and thermal 
excitations. 

It is now well established that at very low temperatures the dynamics of 
Bose-condensed gases can be accurately described by the Gross-Pitaevskii
(GP) equation for the condensate order parameter (see e.g.\ \cite{dalfovo99}).
However, despite considerable work over the past few years, it has proved
far more difficult to extend the theory to include thermal excitations.
One of the major challenges is to capture the dynamics of the non-condensed 
component in a self-consistent way, while still allowing for a feasible 
calculation. A possible approach to this problem was suggested by one of the 
authors (EZ) together with Griffin and Nikuni \cite{nikuni99,zaremba99}. 
In this formalism (which we
will hereafter denote by ZGN) the thermal excitations are described
semiclassically by a Boltzmann kinetic equation, which couples to the
condensate through mean fields and collisions.

In this paper we report on recent progress in solving the ZGN equations to
study the dynamics of Bose-condensed gases in experimentally relevant
regimes. We begin by briefly reviewing the formalism, and present 
analytical calculations of collision rates as a function of temperature. 
We then go on to discuss numerical calculations, where the kinetic equation
is solved using $N$-body simulations with collisions treated through 
Monte-Carlo sampling. Applications to thermal cloud quenching, dipole
modes, and scissors modes \cite{jackson01b} are discussed. We conclude by 
highlighting future research directions.       

\section{ZGN Theory}
The ZGN theory~\cite{zaremba99} is based on the assumption of a Bose 
broken symmetry which
leads to the macroscopic wavefunction $\Phi(\br,t) = \langle \hat
\psi(\br,t)\rangle$. Here, $\hat \psi(\brt)$ is the Bose field operator
and the angular brackets denote an average with respect to a
nonequilibrium density matrix. The condensate density is then given by
$n_c(\brt) = |\Phi(\brt)|^2$. The condensate wavefunction is found to
satisfy the generalized GP equation
\beq
 i\hbar \frac{\partial}{\partial t} \Phi ({\mathbf{r}},t) = \Bigg(
 -\frac{\hbar^2}{2m} \nabla^2 + V ({\mathbf{r}}) + g
 [n_c ({\mathbf{r}},t) + 
 2 \tilde{n} ({\mathbf{r}},t)]-iR ({\mathbf{r}},t) \Bigg) 
 \Phi({\mathbf{r}},t),
\label{eqn1}
\eeq
where $V(\br)$ is the external confining potential, $\tilde n(\brt)$ is
the noncondensate (or thermal atom) density, and $R(\brt)$ is a
non-Hermitian source term that will be defined below. The interactions
between atoms are taken to be local and leads to the mean field 
potential $g[n_c+2\tilde n]$, where 
$g = 4\pi \hbar^2 a/m$, with $a$ the $s$-wave scattering length.
For $T \to 0$, $\tilde n$ and $R$ are zero and one recovers the familiar
nonlinear GP equation. The stationary ground state solution $\Phi_0$
gives the density $n_{c0} = |\Phi_0|^2$ which for large $N$ is
well-approximated by the Thomas-Fermi result $n_{c0} = 
(\mu_0 - V)/g$~\cite{dalfovo99}.
Here, $\mu_0$ is the ground state GP eigenvalue or chemical potential.

Small amplitude excitations of the $T=0$ GP equation can be obtained by
writing $\Phi = \Phi_0 +\delta\Phi$ and linearizing with respect to the
deviation $\delta \Phi$. This procedure is equivalent to the
Hartree-Fock-Bogoliubov (HFB) approximation and generates the Bogoliubov
quasiparticle excitations. At low energies, these excitations can be
interpreted as collective excitations of the condensate, while at high
energies they correspond to single-particle thermal excitations. However,
at nonzero temperatures, one must take into account the occupation of
excited states and the appearance of a noncondensate density $\tilde n$
which interacts with the condensate. This problem was addressed by 
Hutchinson {\it et al.}~\cite{hutchinson97} (hereafter referred to as
HZG) and Dodd {\it et al.}~\cite{dodd98}
who calculated $n_{c0}$ and $\tilde n_0$ self-consistently in the
HFB-Popov approximation. The quasiparticle excitations calculated within
this scheme become temperature dependent and at first sight might be
associated with collective excitations of the trapped gas. However these
excitations in fact correspond to collective excitations of the
condensate in the presence of a {\it static} thermal cloud. In other
words, the dynamics of the latter are not included consistently. This 
was revealed in calculations of the center of mass dipole mode whose
frequency showed a deviation from the trap frequency $\omega_0$,
which is the correct result as dictated by the generalized Kohn 
theorem~\cite{dobson94}.
This was the motivation for developing a theory which consistently
included the dynamics of the thermal cloud with that of the condensate.

The ZGN theory~\cite{zaremba99} provides such a description. It is 
based on the following approximations: (i) the Popov approximation 
which neglects the anomalous average $\tilde m(\brt) =
\langle \tilde \psi(\brt) \tilde \psi(\brt) \rangle$; 
here, $\tilde \psi(\brt) \equiv \hat \psi(\brt) -
\langle \hat \psi(\brt) \rangle$ is the noncondensate field operator,
(ii) the use of Hartree-Fock thermal excitations with energy
$\varepsilon_{\bp}(\brt) = p^2/2m + U(\brt)$, where $U = V+2gn$ is the
mean field potential acting on the thermal atoms, (iii) a semiclassical
approximation for the thermal atoms which are represented by a
phase-space distribution $f(\bprt)$, and (iv) the inclusion of binary
collisions among thermal atoms as well as with the condensate. 

Within these approximations, one obtains the semiclassical 
Boltzmann kinetic equation for the thermal cloud 
\begin{equation} 
 \frac{\partial f}{\partial t} + \frac{{\mathbf{p}}}{m} \cdot \nabla f
 - \nabla U \cdot \nabla_{\mathbf{p}} f 
 = C_{12} [f] + C_{22} [f].
\label{eqn2}
\end{equation}
The collision integrals appearing on the right hand side of this
equation are given by
\bea
&&C_{12}[f]\equiv {\sigma n_c\over \pi m^2} \int d{\bf p}_2\int
d{\bf p}_3 \int d{\bf p}_4 
\delta(m{\bf v}_c+{\bf p}_2-{\bf p}_3-{\bf p}_4)
\delta(\varepsilon_c+\varepsilon_2
-\varepsilon_3-\varepsilon_4) \cr
&&\hskip 2.5truein \times  [\delta({\bf p}-{\bf p}_2)-\delta({\bf
p}-{\bf p}_3)
-\delta({\bf p}-{\bf p}_4)] \cr
&&\hskip 2.5truein \times [(1+f_2)f_3f_4-f_2(1+f_3)(1+f_4)],
\label{eqn3}
\eea
\bea
&&C_{22}[f] \equiv {\sigma \over \pi h^3 m^2}
 \int d{\bf p}_2\int d{\bf p}_3
\int d{\bf p}_4 \delta ({\bf p}+{\bf p}_2 -{\bf p}_3 -{\bf p}_4)
\delta(\varepsilon+\varepsilon_2
-\varepsilon_3-\varepsilon_4) \cr
&&\hskip 2.5truein
\times\left[(1+f)(1+f_2)f_3f_4-ff_2(1+f_3)(1+f_4)\right]\, .
\label{eqn4}
\eea
Here, $\sigma = 8\pi a^2$ is the total cross-section. The $C_{22}$
integral involves the scattering of two atoms from initial to final
thermal states. On the other hand, $C_{12}$ describes collisions in
which a condensate atom is involved in either the incoming or outgoing 
channels.
These collisions have the effect of transferring atoms from or to the
condensate and lead to a change in the number of noncondensate 
atoms, $\tilde N$.  Specifically, we have
\beq
{d\tilde N \over dt} = \int {d\br d\bp \over (2\pi \hbar)^3}\, 
C_{12}[f]\,.
\label{eqn5}
\eeq
These collisions also define the source term
\begin{equation}
 R ({\mathbf{r}},t) = \frac{\hbar}{2n_c} \int \frac{d{\mathbf{p}}}
 {(2\pi\hbar)^3} C_{12} [f]\,
\label{eqn6}
\end{equation}
appearing in Eq.~(\ref{eqn1}). The condensate number $N_c$ of course 
satisfies $dN_c/dt = - d\tilde N/dt$ so
that the total number of atoms, $N = N_c +\tilde N$, is conserved. The
local rates at which atoms enter and leave the condensate are given by
\bea
&&\Gamma_{12}^{in} =
{\sigma n_c\over \pi h^3 m^2} \int d{\bf p}_2\int
d{\bf p}_3 \int d{\bf p}_4 \,
\delta(m{\bf v}_c+{\bf p}_2-{\bf p}_3-{\bf p}_4)
\delta(\varepsilon_c+\varepsilon_2
-\varepsilon_3-\varepsilon_4) \cr
&&\hskip 2.5truein \times (1+f_2)f_3f_4\,.
\label{eqn7}
\eea
\bea
&&\Gamma_{12}^{out} =
{\sigma n_c\over \pi h^3 m^2} \int d{\bf p}_2\int
d{\bf p}_3 \int d{\bf p}_4 \,
\delta(m{\bf v}_c+{\bf p}_2-{\bf p}_3-{\bf p}_4)
\delta(\varepsilon_c+\varepsilon_2
-\varepsilon_3-\varepsilon_4) \cr
&&\hskip 2.5truein \times f_2(1+f_3)(1+f_4)\,.
\label{eqn8}
\eea
In equilibrium, $\bv_c = 0$ and $\varepsilon_c = \mu_{0}$, the GP
eigenvalue. One can then show that these two rates are equal if the
distribution function takes its equilibrium Bose-Einstein form
\beq
f^0(\bp,\br) = {1\over \exp\{ \beta({p^2 /2m}+U_0 -\mu_{0})\} -1}\,,
\label{eqn9}
\eeq
where $\beta = 1/k_B T$.
As would be expected, the condensate and noncondensate must have the
same chemical potential if the net transfer rate is to be zero. However,
for the general nonequilibrium situation the $\Gamma_{12}^{in(out)}$
rates are different and result in a time-dependent amplitude of the
condensate wavefunction. This is revealed most clearly by making use of
phase and amplitude variables, $\Phi(\brt) =
\sqrt{n_c(\brt)}e^{i\theta(\brt)}$. The generalized GP equation is then
equivalent to the continuity equation
\beq
{\partial n_c \over \partial t} + \nabla\cdot(n_c \bv_c) =
\Gamma_{12}^{in} - \Gamma_{12}^{out}\,,
\label{eqn10}
\eeq
where $\bv_c = \hbar \nabla \theta/m$, and the Euler equation
\beq
m {\partial \bv_c \over \partial t} + \nabla\left ( {1\over 2} m v_c^2
\right ) = -\nabla \mu_c
\label{eqn11}
\eeq
with
\beq
\mu_c = -{\hbar^2 \nabla^2 \sqrt{n_c} \over 2m \sqrt{n_c}} + V + gn_c +
2g \tilde n\,.
\label{eqn12}
\eeq

It is instructive to examine $\Gamma^{out}_{12}$ for an equilibrium
state of a typical trapped gas. $\Gamma^{out}_{12}$ represents the
average number of thermal atoms colliding with the condensate per unit
time per unit volume. Thus, dividing $\Gamma^{out}_{12}$ by the local
noncondensate density gives the local equilibrium collision rate per
atom, $1/\tau^0_{12}$, for this kind of collision~\cite{footnote1}. 
In Fig.~1 we plot
$1/\tau^0_{12}$ for $10^5$ Rb atoms in an isotropic trap with
frequency $\nu_0 = 200$ Hz for various temperatures below the ideal gas
critical temperature $T_c^0 \simeq 0.94 \hbar \omega_0 N^{1/3}/k_B$. 
The collision rate has a maximum at the edge of the condensate where the
local fugacity $z(r) = e^{\beta(\mu_{0} -U_0)}$ approaches unity. 
Since
$1/\tau^0_{12}$ is proportional to $n_{c0}$, it falls off rapidly beyond
this point. It is also interesting to note that for $T$ approaching
$T_c^0$, $\omega_0 \tau^0_{12} \le 1$, indicating that a thermal atom in
this region of the condensate has a high probability of making a
$C_{12}$ collision on the timescale of the trap period. However, the
number of thermal atoms which overlap with the condensate is relatively
small and it is therefore more meaningful to define an average $C_{12}$
collision rate per thermal atom by
\beq
{1\over \bar \tau_{12}(T)} = {1\over \tilde N(T)} \int d^3 r {\tilde
n_0(r) \over \tau_{12}^0(r)}\,.
\label{eqn13}
\eeq
This quantity is plotted in Fig.~3 which shows that the thermal cloud is
in the transition regime $\omega_0 \tau_{12}^0 \simeq 1$ rather than the
truly hydrodynamic regime $\omega_0 \tau_{12}^0 \ll 1$ as the local
collision rate in Fig.~1 suggests. Nevertheless, these results show that
$C_{12}$ collisions play an important role in the dynamics of the
thermal cloud in the region of the condensate. Of course, the importance
of these collisions increases with increasing $N$ and one can expect to
be in the hydrodynamic regime for $N \ge 5\times 10^5$ for this
particular trap configuration.

One can also define an equilibrium collision rate between atoms in the
thermal cloud. It is this rate that one would conventionally consider
when deciding between collisionless and hydrodynamic behaviour.
Accordingly, we define a $C_{22}$ collision rate through the relation
\bea
{\tilde n_0 \over \tau_{22}^0} &\equiv& \int {d\bp_1 \over (2\pi \hbar)^3}
C_{22}^{out}[f_1^0]\cr
&=& {\sigma m^6 \over 4\pi h^6} \int d\bv_1 \int d\bv_2 \int d\Omega 
\,v_r f_1^0 f_2^0 (1+f_3^0)(1+f_4^0)\,.
\label{eqn14}
\eea
Here, $C_{22}^{out}$ denotes the term in Eq.~(\ref{eqn4}) in which 
particles in states `1' and `2' scatter into the final states `3' and
`4'. In equilibrium, the $C_{22}^{out}$ and
$C_{22}^{in}$ terms are equal. In Eq.~(\ref{eqn14}),
$\bv_r = \bv_1-\bv_2$ is the relative velocity of the incoming
particles and the solid angle $\Omega$ defines the orientation of the 
final relative velocity vector, $\bv_r' = \bv_3-\bv_4$. Momentum and
energy conservation ensures that $\bv_1+\bv_2 = \bv_3+\bv_4$ and $v_r =
v_r'$. In the classical limit ($h \to 0$), $1/\tau_{22}^0$ reduces to
$\left(1/\tau_{22}^0 \right )_{cl} = \sqrt{2} \sigma v_{th} \tilde n_0$,
where $v_{th} = (8kT/\pi m)^{1/2}$. In Fig.~2 we plot $1/\tau_{22}^0$ 
for the same trap parameters used to generate $1/\tau_{12}^0$. For $T <
T_c^0$, we again see Bose enhancement of the $C_{22}$ collision rate at
the edge of the condensate, with $\omega_0 \tau_{22}^0 \ll 1$ being 
reached in this region at the higher temperatures. In Fig.~3 we present
the average $C_{22}$ collision rate $1/\omega_0\bar \tau_{22}$. 
Below $T_c^0$, this rate is comparable to the average $C_{12}$ collision
rate, indicating that the probability of a thermal atom 
colliding with another thermal atom is similar to the probability of
colliding with the condensate. However, unlike $1/\omega_0\bar\tau_{12}$
which involves the condensate density $n_{c0}$, 
$1/\omega_0\bar\tau_{22}$ 
continues to increase up to $T_c^0$, at which point it reaches a value 
of $\omega_0\bar\tau_{22} \simeq 1$. This limiting value
increases with increasing $N$, indicating that the hydrodynamic regime
is easily reached near $T_c^0$. Above $T_c^0$, the
effect of Bose enhancement rapidly diminishes and the average rate
approaches its classical limiting value.

\section{Numerical Methods}

Eqs.~(\ref{eqn1}) and (\ref{eqn2}) can in principle be used to describe 
the dynamics of a 
trapped Bose gas for arbitrary conditions of temperature, number of
atoms and trap geometry. The equations have been solved previously for
various limiting situations. For example, Eq.~(\ref{eqn1}) was solved 
by HZG~\cite{hutchinson97} for
$R=0$ and with the assumption of a static thermal cloud. As stated
earlier, this neglects the perturbation of the noncondensate 
that arises from the time-dependent mean-field of the oscillating
condensate. As a result, it neglects those modes, such as the
out-of-phase dipole mode, which involve the collective motion of the
thermal atoms. Eqs.~(\ref{eqn1}) and (\ref{eqn2}) have also been used 
to study collective
modes in the collisionless limit using what is effectively a moment
method~\cite{bijlsma99}. 
In this approach, an ansatz for the form of the GP wavefunction and
distribution function of the thermal cloud is made, and a finite number 
of moments of the GP and 
kinetic equations are retained to obtain a closed set of equations for
the modes of interest. However, due to the truncation made,
the effect of Landau damping on the modes is not included. 
Furthermore, the ansatz used for the
form of the nonequilibrium distribution function 
does not treat the mean-field interactions between the two components
accurately. In the opposite limit of high collision rates, we 
previously developed various approximations
which address the dynamics in the hydrodynamic regime~\cite{zaremba99}. 
By assuming
$C_{22}$ collisions to be rapid we developed a description in which the
thermal cloud is assumed to be in local thermodynamic equilibrium, but
not necessarily in local equilibrium with the 
condensate~\cite{nikuni99}. The assumption
of complete local equilibrium results in a Landau-Khalatnikov two-fluid
theory~\cite{khalatnikov65,nikuni01}.
  
Clearly it would be desirable to obtain accurate solutions of our
equations which avoid approximations beyond those 
used to generate the equations themselves. In this way, the limitations
of the approximations underlying the theory can be critically examined.
For this purpose we have developed a numerical procedure within which 
the accuracy of the solutions can be studied and evaluated. We
outline our numerical methods in the following.

The generalized GP equation can be solved by a split-operator technique.
For the purposes of the following discussion, we write Eq.~(\ref{eqn1})
in the operator form
\beq
i\hbar {\partial \Phi \over \partial t} = (T + \tilde V)\Phi\,,
\label{eqn15}
\eeq
where $T$ is the kinetic energy operator and $\tilde V$ is a complex,
time-dependent potential. The wavefunction is advanced in time using the
split-operator method:
\beq
\Phi(t+\Delta t) \simeq e^{-i\tilde V \Delta t/2\hbar} e^{-iT \Delta
t/\hbar} e^{-i\tilde V \Delta t/2\hbar} \Phi(t)\,.
\label{eqn16}
\eeq
Since the $\tilde V$ operator is local in position space the effect of
$e^{-i\tilde V \Delta t/2\hbar}$ is most easily treated in the
coordinate representation. The kinetic energy operator on the other hand
is local in momentum space and it is therefore useful to Fourier
transform the intermediate state when determining the effect of $e^{-iT
\Delta t/\hbar}$. A fast fourier transform (FFT) routine is used for 
this purpose. The inverse transform is then applied in order to complete
the evaluation of the evolution operator. Since $\tilde V$ is
time-dependent, a higher order approximant to the evolution operator is
in fact used to increase the accuracy and stability of the GP equation
solution. 

The use of the semiclassical Boltzmann equation for the thermal atoms
corresponds to a Hamiltonian description of the particle dynamics. In
our case the Hamiltonian is simply $H_{cl} = p^2/2m + U(\brt)$, where
$U(\brt) = V(\br) + 2gn(\brt)$ is the time-dependent mean-field
potential. Thus the trajectory of a given atom between collisions can be
determined using Newton's equations of motion. The evolution of the
distribution function $f(\bprt)$ is then determined by considering the
dynamics of $N_T$ representative test
particles~\cite{bird94,wu97,lopez-arias98} which in effect
corresponds to the distribution function $f_T(\bprt) = \sum_{i=1}^{N_T}
\delta(\br -\br_i) \delta(\bp-\bp_i)$. In the absence of collisions, the
dynamics of this swarm of test particles is equivalent to solving the
collisionless Boltzmann equation.

In reality, the deterministic evolution of a particular phase space 
trajectory is interrupted stochastically by collisions with other atoms
or with the condensate. In other words, in a time step $\Delta t$, a
given thermal atom has some probability of suffering a collision, and 
the numerical algorithm used to generate these probabilities must yield
collision rates that are consistent with the collision integrals in the
Boltzmann equation. To simulate the $C_{22}$
collisions~\cite{jackson01a}, we bin the test
particles into position space volume elements $\Delta^3 r$. Pairs of
atoms ($ij$) chosen at random from a given volume element define
collision partners, and the probability of their suffering a collision 
in a time interval $\Delta t$ is given by
\beq
P_{ij} = \tilde n \sigma |\bv_i-\bv_j|\int {d\Omega \over 4\pi}
(1+f_3)(1+f_4)\Delta t\,.
\label{eqn17}
\eeq
The angular integral is an average over all possible orientations of the
final relative velocity $\bv_r'$ and can be replaced by a single
scattering event by choosing a random solid angle $\Omega_R$ uniformly
distributed on a unit sphere. Thus the probability used in the
simulation is
\beq
P_{ij} = \tilde n \sigma |\bv_i-\bv_j|
(1+f_3^{\Omega_R})(1+f_4^{\Omega_R})\Delta t\,.
\label{eqn18}
\eeq
The final velocities $\bv_3$ and $\bv_4$ are determined by $\bv_i$,
$\bv_j$ and $\Omega_R$ through momentum and energy conservation.
To test whether a collision actually occurs, a random number $R_{ij}$
distributed uniformly between 0 and 1 is chosen and compared to 
$P_{ij}$. If $R_{ij}$ is less than $P_{ij}$, the collision is accepted.
This process is repeated for each pair ($ij$) in every real-space volume
element. We have checked that the procedure correctly reproduces
the equilibrium collision rates as given by Eq.~(\ref{eqn14}).

The $C_{12}$ collisions are treated in a similar fashion. In this case,
both `in' and `out' collisions, as illustrated schematically in Fig.~4,
must be considered separately since one of the incoming or outgoing
particles is from the condensate. The overall rate of `out' collisions
is given by Eq.~(\ref{eqn8}), in which a particle with velocity $\bv_2$ 
collides
with the condensate to produce two outgoing thermal atoms with
velocities $\bv_3$ and $\bv_4$. This process is similar to a $C_{22}$
collision in that the final relative velocity $\bv_r' = \bv_3 -\bv_4$ 
can have arbitrary direction with a magnitude of $v_r' =
\sqrt{|\bv_c-\bv_2|^2 -4gn_c/m}$. This value is a consequence of energy
conservation and accounts for the fact that the mean-field energy of a
thermal atom differs from that of a condensate atom by $gn_c$. 
Choosing one of the test particles, $i$, from the volume element 
$\Delta^3r$, the probability of it suffering an `out' collision is 
given by
\beq
P_i^{out} =  n_c \sigma v_r'
(1+f_3^{\Omega_R})(1+f_4^{\Omega_R})\Delta t\,.
\label{eqn19}
\eeq

The situation for `in' collisions is somewhat more complex since an
arbitrary pair of thermal atoms in $\Delta^3r$ will be precluded by
energy and momentum conservation from scattering into the condensate.
To ensure that these collision events are sampled correctly, we rewrite
Eq.~(\ref{eqn7}) by interchanging the `2' and `3' labels in the 
integral. The
`2'-particle is then paired with the `4'-particle to produce the
outgoing `3'-particle. The resulting integral
leads to a significantly different form for the probability of this
type of collision event. Without giving details, we find that
($i =$ `2')
\beq
P_i^{in} =  {n_c \sigma {\cal A}_v \over \pi |\bv_i -\bv_c|} 
(1+ f_3^{\bv_R}) f_4^{\bv_R}\Delta t\,,
\label{eqn20}
\eeq
where $\bv_R$ is a random point in a plane of area ${\cal A}_v$ in 
velocity space. The plane is defined by energy and momentum
conservation, and once the velocity $\bv_R$ is chosen, the velocities
$\bv_3$ and $\bv_4$ are determined. $P_i^{in}$ is largely independent of
${\cal A}_v$ as long as the plane completely samples the occupied 
regions of phase space. We have checked that $P_i^{out}$ and
$P_i^{in}$ give identical collision rates for the equilibrium
situation. The simulation of $C_{12}$ collisions then proceeds as
follows. For each particle $i$ in $\Delta^3r$, $P_i^{out}$ and
$P_i^{in}$ are calculated. A random number $R_i$ is selected and if $R_i
> P_i^{in}+P_i^{out}$, no collision occurs, while if $R_i < P_i^{in}$,
the `in' collision is accepted and if $P_i^{in} < R_i <
P_i^{in}+P_i^{out}$, the `out' collision is accepted. This procedure
ensures that any given particle only makes one kind of collision, if
a collision occurs at all. We have again checked that this prescription
gives correctly the equilibrium `in' and `out' collision rates.

We finally address the question of determining the mean-field potential,
$2g \tilde n$, for $N_T$ point-like test particles. An approximation to
$\tilde n(\br)$ is obtained by dividing space into cells $\Delta^3 r$
and binning the distribution to construct a density histogram. 
If $\Delta^3 r$ is sufficiently small and $N_T$ sufficiently large, this
procedure will generate a fairly smooth density distribution.
Since our simulations are restricted practically to $N_T < 10^6$ and
a three-dimensional spatial grid that is relatively coarse,
statistical fluctuations in the number histogram are inevitable. To
compensate for these fluctuations we have adopted the following
strategy. First, we use a cloud-in-cell method~\cite{hockney81}
to define density 
weights at the 8 grid points bounding a cell according to the position 
of a particle within the cell. Having defined a density on the grid
points of our mesh, we next convolve the density with a gaussian:
\beq
\tilde n_{conv}(\br) = {1\over \pi^{3/2}\eta^3}
\int d\br'  e^{-|\br-\br'|^2/\eta^2} \tilde n(\br')\,.
\label{eqn21}
\eeq
By choosing a width parameter $\eta$ somewhat larger than the spatial
grid used in our calculations, we achieve a further smoothening of the
thermal atom density. The final mean-field potential of the thermal
atoms is then defined as $2g\tilde n_{conv}(\br)$.
This result can also be viewed in terms of an interaction potential
$g(\br-\br') = (g/\pi^{3/2}\eta^3) \exp(-|\br-\br'|^2/\eta^2)$ 
which has a finite range $\eta$ as opposed
to the contact interaction $g\delta(\br-\br')$ usually assumed. Such a
procedure has a minor effect for density variations on length scales
larger than $\eta$ but has the benefit of reducing fluctuations 
stochastically generated on the scale of the spatial grid.
We can of course check that the physical properties of interest are
unaffected by the value of $\eta$ used in the calculations. The gaussian
convolution has the further useful feature of being readily
calculable with a FFT routine.

Our simulations commence with the specification of the initial state of
the system, namely the condensate wavefunction $\Phi(\br,0)$ and the
initial
distribution of test particles. The particular choice is dictated by the
physical situation being treated and several examples are considered in
the following section. Once the initial state is known, the condensate
and noncondensate mean-field potentials are determined. The condensate
wavefunction is propagated for a time $\Delta t$ and the test particle
positions and momenta are updated. At the end of this time step, the
$C_{22}$ and $C_{12}$ collisions are considered in turn and their net
effect on the thermal cloud distribution is determined. At the same
time, the $C_{12}$ collisions are accumulated to construct the source
term $R(\br,t)$ that is needed for the next cycle in the GP evolution.
This procedure is then repeated for the duration of the simulation.
Quantities of physical interest can be calculated from the condensate
wavefunction and test particle distribution at each time step and are
recorded for subsequent analysis and display.

\section{Applications}
\subsection{Equilibration}
We begin by considering a relatively simple model problem which
illustrates the equilibration of an initial nonequilibrium state. We
imagine a trapped gas in equilibrium at some temperature $T_0$ which is
chosen such that approximately half of the atoms are in the condensate.
To be specific, we consider $5\times 10^4$ $^{87}$Rb atoms in an 
isotropic trap with
harmonic frequency $\nu_0 = 187$ Hz. The scattering length is $a = 5.82$
nm.
The equilibrium state is determined by solving self-consistently the
following pair of equations:
\beq
-{\hbar^2 \nabla^2 \over 2m} \Phi_0(\br) + [V(\br) + gn_{c0}(\br) + 
2g \tilde n_0(\br) ] \Phi_0(\br) = \mu_0 \Phi_0(\br)\,,
\label{eqn22}
\eeq
and
\beq
\tilde n_0(\br) = {1\over \Lambda^3} g_{3/2}(z(\br))\,,
\label{eqn23}
\eeq
where $\Lambda = (2\pi \hbar^2/mkT_0)^{1/2}$ is the thermal de Broglie
wavelength, $z_0(\br) = \exp(\beta_0(\mu_0 - U_0(\br))$ is the local
equilibrium fugacity, $\beta_0 = 1/k_B T_0$ and 
$U_0(\br) = V(\br) + 2g[n_{c0}(\br) + \tilde n_0(\br)]$. For $T_0=200$ 
nK, we find $N_{c0} = 2.58 \times 10^4$. We next define a nonequilibrium
state by first constructing a sample of test particles representative of
the equilibirium density $\tilde n_0(r)$. Each of these particles is
then distributed in momentum according to the 
phase space distribution $f(\bp,\br) = (\Lambda/\Lambda_0)^3
[z_0(\br)^{-1}\exp(\beta p^2/2m) - 1]^{-1}$, where the temperature $T$
is half the equilibrium value $T_0$. This initial nonequilibrium state
has a total density equal to the equilibrium distribution, but the
cloud is too `cold' to sustain this distribution as a function of time.
It begins to collapse in real space, with $C_{12}$ collisions
simultaneously transferring atoms from the thermal cloud to the
condensate. In Fig.~5 we show the evolution of the condensate and
noncondensate numbers as a function of time.
One can see that the condensate number relaxes to a higher
value corresponding to a final equilibrium temperature $T' < T_0$, on a
timescale expected on the basis of the equilibrium $C_{12}$ collision
rates. The figure also shows the total number of atoms, $N=N_C+\tilde
N$, as a function of time in the simulation. This number should of
course be constant but the simulation does not conserve particle
number exactly at each time step 
because of the different ways in which particle number
is changed in the condensate and thermal cloud. The Monte-Carlo sampling
of collisions leads to integer-valued changes in the number of thermal 
test particles. On the other hand, the normalization of the condensate
changes quasicontinuously at a rate determined by the strength of the
source term $R(\brt)$ in the GP equation. The fact that $N$ in our 
simulations exhibits only small fluctuations about the initial value
is reassuring and demonstrates that
number conservation is preserved in an average sense. This is an
obvious requirement of any method that hopes to address problems
such as condensate growth~\cite{bijlsma00}.

In Fig.~6 we show the root-mean-squared radius
$\sqrt{\langle r^2 \rangle}$ of both the condensate and thermal cloud.
The thermal cloud initially begins to collapse as a result of the quench
before bouncing back and oscillating at a frequency very close to the
monopole frequency $2\omega_0$ of a noninteracting thermal cloud.
Oscillations in the condensate are in turn 
induced by the time-dependent mean-field potential as well as by the
condensation of thermal atoms. These oscillations at first build up in
time and then damp out as the entire system approaches overall
equilibrium. The frequency of these oscillations is about 5\% lower than
the TF frequency for the condensate monopole mode, 
$\sqrt{5}\omega_0$~\cite{stringari96a}. This
confirms that our simulations are providing a good description of
the dynamical behaviour of both the condensate and thermal cloud.

\subsection{Dipole Oscillations}
Our second example is perhaps an even more stringent test of our
numerical procedures. For harmonic confinement, a trapped gas must
exhibit a centre of mass oscillation in which the total density of the 
gas oscillates rigidly without change in shape at the trap frequency
$\omega_0$, i.e., $n(\brt) = n_0(\br-\bbox{\eta}(t))$, with 
$\bbox{\eta}(t) = \bbox{\eta}_0 \cos(\omega_0 t)$.
This result is a consequence of the generalized Kohn
theorem~\cite{dobson94}
and must be satisfied in any consistent theory of the dynamics. The
failure to satisfy this property was the signature that the HFBP mode
calculations were deficient~\cite{hutchinson97}. The ZGN theory, 
however, is formally consistent with the generalized Kohn
theorem~\cite{zaremba99}. Our purpose here
is to demonstrate that our numerical implementation of the theory 
satisfies this requirement as well.

In these simulations, we retain the same trap parameters as given in the
previous subsection and the same number of Rb atoms, but at a higher
temperature ($T_0 = 225$ nK)
where the difference between the condensate and noncondensate numbers is
larger. The reason for wanting this difference to be appreciable will 
become apparent shortly. We begin
by obtaining the equilibrium condensate wavefunction and an equilibrium
distribution of test particles. To excite the dipole oscillations, we
then displace each component a small distance $\Delta z$ along the 
$z$-axis relative to the trap, but in opposite directions. This is 
easily achieved by
shifting the condensate wavefunction a certain number of grid points on
the spatial mesh, and adding a constant displacement $\Delta z$ to the
position of each test particle. The two components are then released and
their subsequent motion is determined.
 
In our first simulation, we turn off all collisional processes. The two
components then interact only as a result of the mean-field potentials.
The mean displacement of the condensate, $\langle z_c(t)\rangle = \int
d\br zn_c(\br,t)/N_c$, and the thermal cloud, $\langle \tilde z(t) 
\rangle = \sum_{i=1}^{N_T} z_i(t)/N_T$, are then
calculated as a function of time. In Fig.~7 we plot the center of mass
position
\beq
Z_{cm}(t) = {1\over N} \left ( N_c \langle z_c(t) \rangle + \tilde N
\langle \tilde z(t) \rangle \right )\,,
\label{eqn24}
\eeq 
and the relative displacement
\beq
z(t) = \langle z_c(t) \rangle - \langle \tilde z(t) \rangle \,.
\label{eqn25}
\eeq
We see that $Z_{cm}(t)$ oscillates at the trap frequency, $\omega_0$, as
expected, with no change in amplitude. This amplitude is given by
$Z_{cm}(0) = \Delta z (\tilde N - N_C)/N$ and is non-zero because $\tilde
N \ne N_c$ in the equilibrium state. The relative displacement $z(t)$ is
also shown, and it is seen to decay as a function of time, although its
behaviour at longer times is more complex.

These results can be interpreted in terms of a two mode
model~\cite{zaremba99}: (i) a
centre-of-mass mode for which $\langle z_c(t) \rangle_1 /\langle 
\tilde z(t)
\rangle_1 =1$ and (i) an out-of-phase dipole mode for which $\langle
z_c(t) \rangle_2 / \langle \tilde z(t) \rangle_2 = -\tilde N/N_c$.
Within this picture, the initial displacement of the two components
excites a superposition of these two modes and the individual
displacements are given by
{\setlength{\jot}{10pt}
\bea
\langle z_c(t) \rangle &=& \left [\left ((\tilde N-N_c) / N\right ) 
\cos
\omega_1 t - (2\tilde N / N) \cos \omega_2 t \right ] \Delta z\,,
\nonumber \\
\langle \tilde z(t)\rangle &=& \left [ \left ( (\tilde N - N_c) / N)
 \right ) \cos
\omega_1 t + (2N_c /N) \cos \omega_2 t \right ] \Delta z \,,
\label{eqn26}
\eea
}
where $\omega_1$ and $\omega_2$ are the two mode frequencies. In this
model, $Z_{cm}(t) = \Delta z (\tilde N -N_c)/N \cos \omega_1 t$ and
$z(t) = - \Delta z (4\tilde N/N) \cos \omega_2 t$. These expressions
qualitatively describe the results of the full simulation, although the
relative displacement is actually damped as a result of coupling to
other internal degrees of freedom. As the condensate moves through the
thermal cloud, the thermal atoms experience a time-dependent mean-field
potential which can transfer energy to them. This excitation mechanism 
is the source of Landau damping of the mode.
However, the interplay between the internal degrees of freedom
and the out-of-phase collective variable is complex and is the reason
why $z(t)$ does not exhibit a simple damped-exponential behaviour.

In our second dipole simulation, we include both the $C_{12}$ and
$C_{22}$ collision processes. The qualitative behaviour is similar
to our previous simulation in that the centre-of-mass mode continues to
oscillate without damping at the trap frequency, $\omega_0$. This is the
signature that the generalized Kohn theorem is being satisfied. In
contrast to the centre-of-mass mode, the out-of-phase dipole mode is
damped. Interestingly, the behaviour is more regular than in the 
absence of collisions. The damping is still due to the Landau mechanism,
but as these mean-field excitations proceed, collisions are
continually driving the gas towards local equilibrium. These collisions
limit the extent to which the thermal cloud phase space distribution can
be driven out of equilibrium, and as a result, the complex long-time
behaviour seen in the absence of collisions is eliminated.

\subsection{Scissors Mode Oscillations}

As our final application, we consider the scissors mode oscillations
that have recently been studied in anisotropic
traps~\cite{marago00,marago01}. A preliminary
report of our work has appeared~\cite{jackson01b} and we shall here 
summarize some of the main findings.

We consider an anisotropic harmonic trap of the form $V(\br) =
m(\omega_\rho^2\rho^2 + \omega_z^2 z^2)/2$ which has axial symmetry
about the $z$-axis, with $\omega_z \simeq \sqrt{8} \omega_\rho$. In the
experiments~\cite{marago01}, 
the scissors mode is excited by adiabatically rotating the
trap potential through a small angle $\theta_0$ and then suddenly
rotating the trap through an angle $-2\theta_0$. To simulate this
process, we first determine the equilibrium state of the system at some
temperature $T$ using the method outlined earlier. We then rotate the
particle coordinates and condensate wavefunction relative to the
trapping potential through an angle $2\theta_0$ about the $y$-axis.
At this point the dynamical simulations commence.

The relevant dynamical variables in this situation are the
$xz$-quadrupole moments $Q_c(t) = \int d\br\,xz n_c(\brt)$ and $\tilde
Q(t) = (\tilde N/N_T)\sum_{i=1}^{N_T} x_i(t) z_i(t)$. If we assume that
the two components rotate about the $y$-axis without distortion, these
quadrupole moments would be given by $Q_\alpha = \theta(t) Q_\alpha^0$,
where $\theta(t)$ is the instantaneous rotation angle and $Q_\alpha^0 =
\langle x^2-z^2\rangle_\alpha^0$ is the equilibrium quadrupole moment of
the $\alpha$-th component. We thus define the `rotation angle'
$\theta_\alpha(t) \equiv Q_\alpha(t)/Q_\alpha^0$ as
the variable representing the angular displacement of the two 
components in our simulations. The advantage of using the angular 
variables is that the range of angular displacements is similar for 
both components, and independent of temperature, in contrast to the
quadrupole moments.

In Figs.~8-10 we show the angular displacements of the two components 
as a
function of time for a few temperatures. At the intermediate temperature
of $T= 185$ nK, $N_c/N = 0.33$, and we see the condensate exhibiting a
damped oscillation at a single frequency, while the thermal cloud
oscillation consists of a superposition of two frequencies. This is to
be expected on the basis of the modes that are found at low temperatures
for a pure condensate and for a thermal cloud above
$T_c^0$~\cite{guery-odelin99}.
For the
$T= 0$ condensate, a straightforward analysis of 
the linearized TF equations of motion which follow from 
Eqs.~(\ref{eqn10})-(\ref{eqn12}) allows one to determine the scissors 
mode frequency. 
Multiplying the continuity equation by $xz$
and integrating over all space leads to
\beq
{d\over dt}\langle xz\rangle = \langle xv_{cz} + zv_{cx}\rangle \,,
\label{eqn27}
\eeq
where the velocity moments are defined as $\langle x_i v_{cj} \rangle =
\int d\br\,x_i v_{cj}(\brt) n_{c0}(\br)$. Taking these moments of the
velocity equation gives
\beq
{d\over dt}\langle xv_{cz} + zv_{cx}\rangle=
-(\omega_x^2+\omega_z^2) \langle xz\rangle\,.
\label{eqn28}
\eeq
These equations define the condensate scissors mode with frequency
$\omega_{sc} = \sqrt{\omega_x^2+\omega_z^2}$. Our $T=0$ simulations for 
$N= 2\times 10^4$ yield a frequency which is 1.5\% larger than the TF
result due to finite number effects.

A similar moment analysis can be carried out for the Boltzmann equation
above $T_c^0$~\cite{guery-odelin99}. One obtains the following moment 
equations 
\bea
{d\over dt}\langle xz\rangle &=& \langle xv_z+zv_x\rangle \nonumber \\
{d\over dt}\langle xv_z - zv_x\rangle&=& 2 \epsilon \omega_0^2 
\langle xz\rangle \nonumber \\
{d\over dt}\langle xv_z + zv_x\rangle&=& 2\langle v_x v_z\rangle
-2\omega_0^2 \langle xz \rangle \nonumber \\
{d\over dt}\langle v_xv_z\rangle&=&\ -\omega_0^2 \langle x v_z + z v_x
\rangle -\epsilon \omega_0^2 \langle xv_z - zv_x\rangle -
{\langle v_x v_z\rangle \over \tau} \,,
\label{eqn29}
\eea
where $\epsilon = (\omega_z^2-\omega_x^2)/(\omega_z^2+\omega_x^2)$ and
$\omega_0^2 = (\omega_x^2+\omega_z^2)/2$.
In the collisionless limit ($\tau \to \infty$), these equations yield
two scissors modes with frequencies $\omega_\pm = \omega_z\pm
\omega_\rho$. Our simulations above $T_c^0$ agree well with these
predictions. As Figs.~8-10 indicate, these modes persist down to quite 
low temperatures below $T_c^0$ due
to the relatively weak mean-field interactions between the condensate
and thermal cloud.

We now discuss the temperature dependence of the scissors mode
oscillations. Fig.~8 shows results for $T=55$ nK and $N_c/N = 0.95$. 
The condensate mode is only weakly damped
with a frequency which is close to the $T=0$ limit. However, the
dynamics of the thermal cloud is quite different from the behaviour seen
at higher temperatures. At this low temperature, the
dynamics of the thermal cloud is strongly affected by the condensate 
mean-field. The fact that the thermal cloud shows an oscillation at the
condensate frequency is indicative of its being driven by the more
massive condensate. With increasing temperature, this coupling
weakens and the thermal cloud begins to display its own scissors mode 
frequencies (Fig.~9). At the same time, the damping of the
condensate mode is seen to increase as a result of the increasing rate
of Landau damping. Finally, at the highest temperature of 296 nK shown
in Fig.~10, $N_c/N
= 0.02$, and the roles of condensate and thermal clould are reversed in
that the condensate is strongly coupled to the more massive thermal
cloud and exhibits a more complex oscillation pattern.

The interaction between the condensate and thermal cloud in the course
of their oscillations is revealed by a density contour plot for the two
components. This is shown in Fig.~11 for a sequence of equally spaced
time steps spanning approximately half a period of the condensate
oscillation. What is striking about this figure is the lack of any
obvious distortion of the condensate. It appears very much to be a rigid
body (but not in a rotational sense!) as it oscillates in the
anisotropic trap. The thermal component on the other hand shows clear
signs of its interaction with the repulsive mean-field of the
condensate. The condensate tends to move the thermal cloud out of its
way in the course of its motion, much like a boat moving through water.
Clearly the thermal cloud is strongly perturbed in the vicinity of the
condensate. It is here that excitation of the thermal atoms, which leads
to Landau damping of the condensate oscillation, occurs most strongly.

A comparison of our results with experiment is shown in Fig.~12.
To obtain
the frequencies and damping rates of the various scissors modes we make
the assumption that the calculated time-dependent quadrupole moments 
are a superposition of three normal modes, one corresponding to the
condensate scissors mode and the other two to the thermal cloud scissors
modes. Accordingly, we fit the time-dependent data in Figs.~8-10 with 
three
damped sinusoidal functions with adjustable amplitudes.
The values of the frequency and damping rate of the condensate scissors
mode extracted in this way are found to be in very good agreement with
experiment for $T < 0.8 T_c^0$. Discrepancies occur above this
temperature, but here the experimental error bars are quite large. As
discussed above, the condensate is strongly coupled to the thermal cloud
as $T_c^0$ is approached and it becomes difficult to obtain good fits to
the condensate quadrupole moment using the three-mode analysis. This 
is probably also the source of the large error bars associated with the 
experimental fits. 

At low temperatures, the frequency of the condensate mode at first
increases slightly with increasing temperature. This we believe is due
to an increasing deviation from the TF frequency $\sqrt{\omega_\rho^2 +
\omega_z^2}$ as a result of the decreasing number of condensate atoms.
At higher temperatures, the frequency begins to decrease as a result of
the increasing size of the thermal cloud. As the density contours in
Fig.~11 suggest, the condensate drags part of the thermal cloud along 
with it and one would expect the condensate to effectively have a
larger inertia, and therefore a lower frequency. The damping of the
condensate mode is essentially due to Landau damping but $C_{12}$ and
$C_{22}$ collisions are indirectly playing an important role. Without
collisions, the damping rate is as much as 50\% lower than the values
shown in the figure. Collisions have the effect of equilibrating the
thermal cloud which is being driven by the condensate, and it appears
that this internal equilibration is enhancing the rate at which 
excitations of thermal atoms occur.

The existence of a single condensate scissors mode is a conseqence of
the irrotational nature of the condensate velocity field. In contrast,
the velocity fields of each of the two thermal cloud modes have both
irrotational and solenoidal components. This distinction becomes
apparent in the moment of inertia of the system, $I$. Zambelli and
Stringari~\cite{zambelli01} have shown that the moment of inertia is 
given by the expression
\beq
\frac{I}{I_{\rm rigid}} = (\omega_z^2 - \omega_\rho^2)^2 
\frac{\int {\rm d}\omega \, \chi''(\omega) / \omega^3}{\int 
{\rm d}\omega \, \chi'' (\omega) \, \omega}, 
\label{eqn30}
\eeq
where $I_{\rm rigid}$ is the moment inertia of the system treated as a
rigid body. The quantity $\chi'' (\omega)$ is the imaginary part of the
quadrupole response function of the system:
\beq
\chi (\omega) =  {i\over \hbar}\int_0^\infty dt \langle [
\hat Q(t), \hat Q(0) ]\rangle_0\,,
\label{eqn31}
\eeq
where $\hat Q = \sum_{i=1}^N \hat x_i \hat z_i$ is the quadrupole moment
operator. This function determines the linear response of the system to
an arbitrary time-dependent perturbation that couples to $\hat Q$. In
particular, the sudden rotation of the trap through a small angle
$\theta_0$
induces the time-dependent quadrupole moment $Q(t)$ determined in our
simulations. It can be shown that
\beq
\chi''(\omega) = -\omega \Re \left \{ \tilde Q'(\omega)/\lambda
\right \}\,,
\label{eqn32}
\eeq
where $\tilde Q'(\omega)$ is the Fourier transform of
$Q(t)-Q(\infty)$ and $\lambda = m(\omega_z^2 - \omega_x^2) \theta_0$.
Our three-mode analysis of the induced quadrupole moment yields a
$\chi''(\omega)$ consisting of a sum of Lorentzian spectral densities
for which the frequency moments required in Eq.~(\ref{eqn30}) are 
undefined. However, the moment of inertia can be estimated by replacing
the Lorentzian line shapes by delta functions having the same spectral
weight. Using this procedure, we show the moment of inertia of the 
system as a function of temperature in Fig.~13. At high temperatures, 
$I$ tends towards $I_{\rm rigid}=m \langle x^2+z^2 \rangle N$, as 
expected when the large thermal cloud dominates the response.
On the other hand, as $T \rightarrow 0$, the moment of inertia takes on
the irrotational value, $I=\bar \epsilon^2 I_{\rm rigid}$, where
$\bar \epsilon=\langle x^2-z^2
\rangle_c^0 / \langle x^2+z^2 \rangle_c^0$ parameterises the deformation
of the condensate in equilibrium. 

Under the semiclassical approximation, Stringari~\cite{stringari96b} 
showed (at least in the ideal gas case) that $I$ at
intermediate temperatures is simply the sum of the two limiting values,
weighted by the number of atoms in each component
\begin{equation}
 I = \bar \epsilon^2 m \langle x^2 + z^2 \rangle_c N_c + m \langle
 x^2+z^2 \rangle_n \tilde N\,.
\label{eqn33}
\end{equation}
This quantity can be readily calculated from the equilibrium densities
and is plotted (as a ratio with $I_{\rm rigid}$) in Fig.~\ref{inertia}.
Comparison to the results from (\ref{eqn30}) show that both 
monotonically increase
with rising temperature, clearly indicating the transition between
superfluid and normal fluid behaviour. Small discrepancies between the
two approaches may result from our evaluation of the frequency spectrum,
or in generalising (\ref{eqn33}) to an interacting gas.

\section{Conclusions}
We have used the ZGN theory to investigate the dynamics of trapped Bose
gases at finite temperatures. In this theory, the condensate is
described by a generalized GP equation while the thermal cloud is
treated semiclassically in terms of a Boltzmann kinetic equation. We
have developed a numerical procedure for solving these equations and
have applied the method to a variety of dynamical problems including
equilibration and collective excitations. We have demonstrated that the
ZGN theory provides an effective method for dealing with such problems 
and is a useful tool in the interpretation of experiment. In particular,
the method has been used to examine the scissors modes in anisotropic
traps. The good agreement with experiment found for this entire
class of modes is encouraging and suggests that other situations can
also be analyzed successfully. We plan to continue this work with a
systematic study of the various collective modes that have been probed
experimentally, as well as problems associated with vortex nucleation
and dynamics.

\acknowledgements
We acknowledge financial support from the Natural Sciences and
Engineering Research Council of Canada, and
the use of the HPCVL computing facilities at Queen's. We thank A. 
Griffin, J. Williams and T. Nikuni for valuable discussions.

\begin{figure}
\centering
\psfig{file=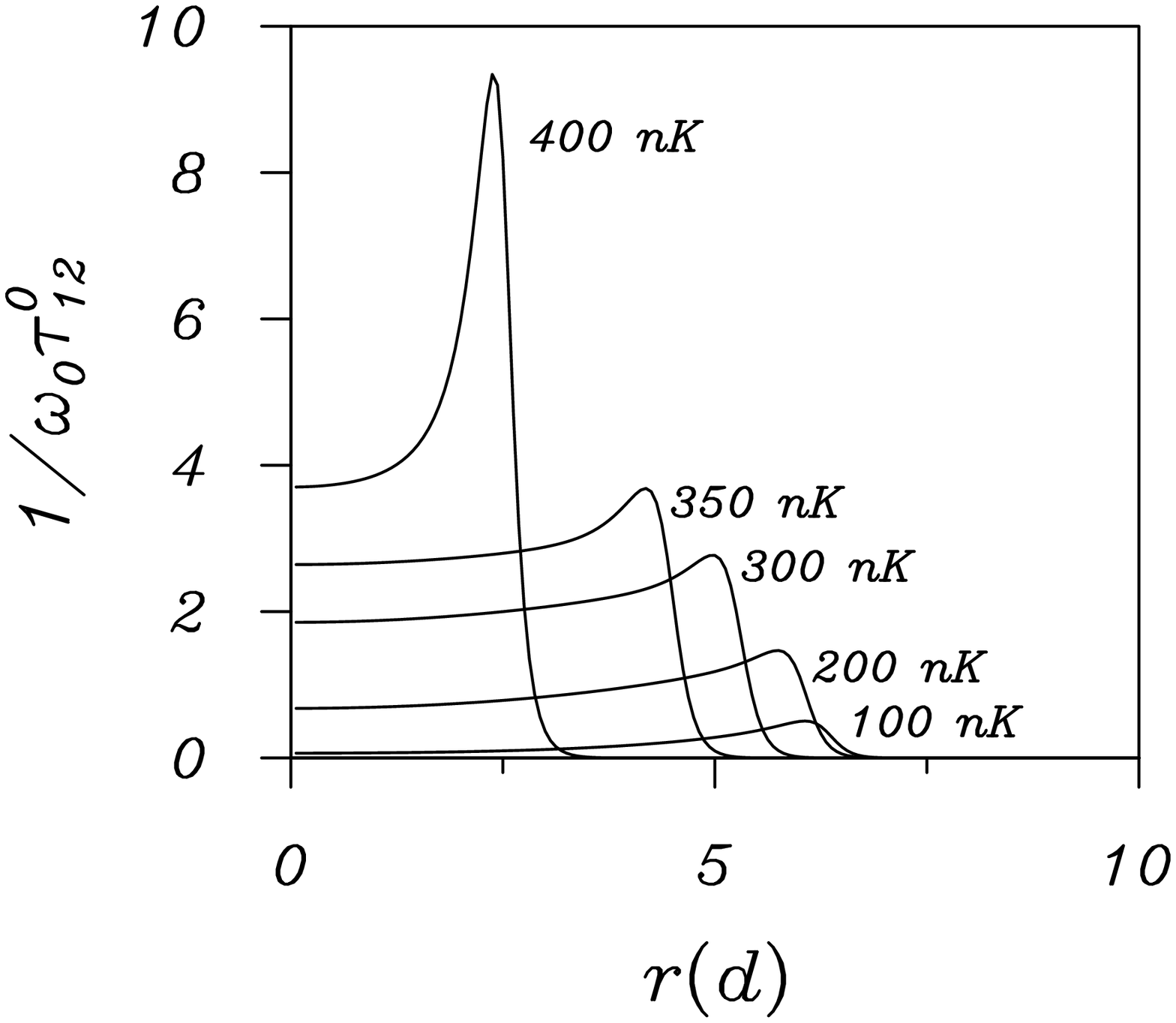, scale=0.55, angle=0, bbllx=25, bblly=140, 
 bburx=540, bbury=620}
\caption{\label{tau_12}
The equilibrium local $C_{12}$ collision rate $1/\tau_{12}^0$ in units
of the trap frequency, $\omega_0$, as a function of the radial distance 
in units of the harmonic oscillator length $d = \sqrt{\hbar/m\omega_0}$.
The different curves are labelled by the temperature $T$. The
trap is isotropic with harmonic frequency $\nu_0 = 200$ Hz and contains
$10^5$ $^{87}$Rb atoms.
 }
\end{figure}

\begin{figure}
\centering
\psfig{file=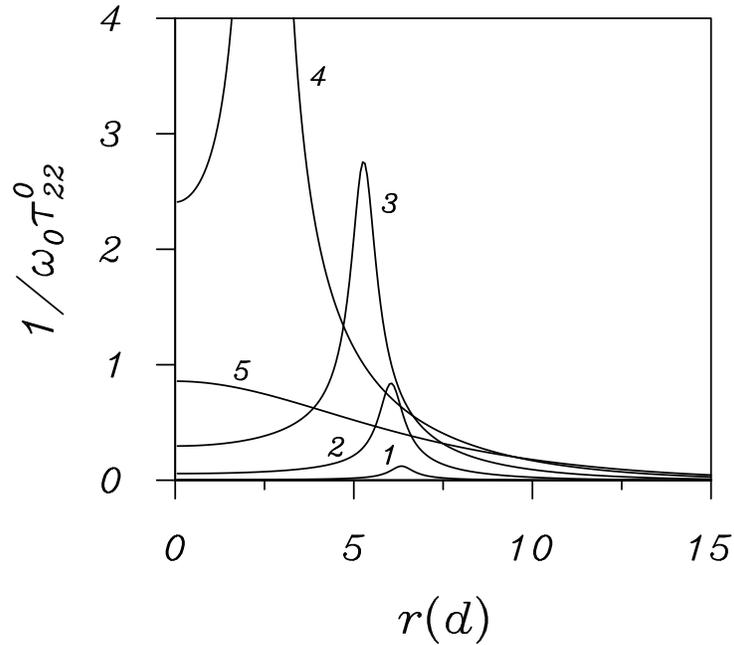, scale=0.55, angle=0, bbllx=35, bblly=140, 
 bburx=540, bbury=610}
\caption{\label{tau_22} 
The equilibrium local $C_{22}$ collision rate $1/\tau_{22}^0$ plotted as
$1/\tau_{12}^0$ in Fig.~1. The curves labelled 1 through 5 correspond to
temperatures between 100 nK and 500 nK in steps of 100 nK.
}
\end{figure}

\begin{figure}
\centering
\psfig{file=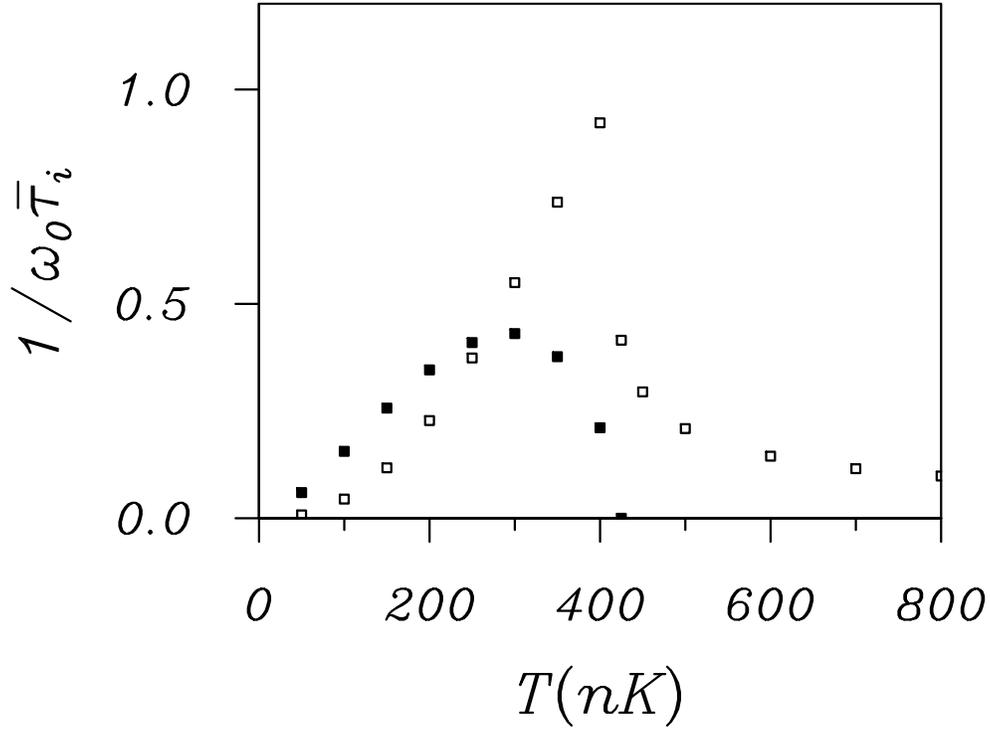, scale=0.7, angle=0, bbllx=15, bblly=140, 
 bburx=550, bbury=570}
\caption{\label{avgtau}
Average collision rates per atom in units of the trap frequency,
$\omega_0$, as a function of temperature. 
The solid squares are for $C_{12}$ collisions and the open
squares are for $C_{22}$ collisions. The trap parameters are the same as
in Fig.~1.
 }
\end{figure}

\begin{figure}
\centering
\psfig{file=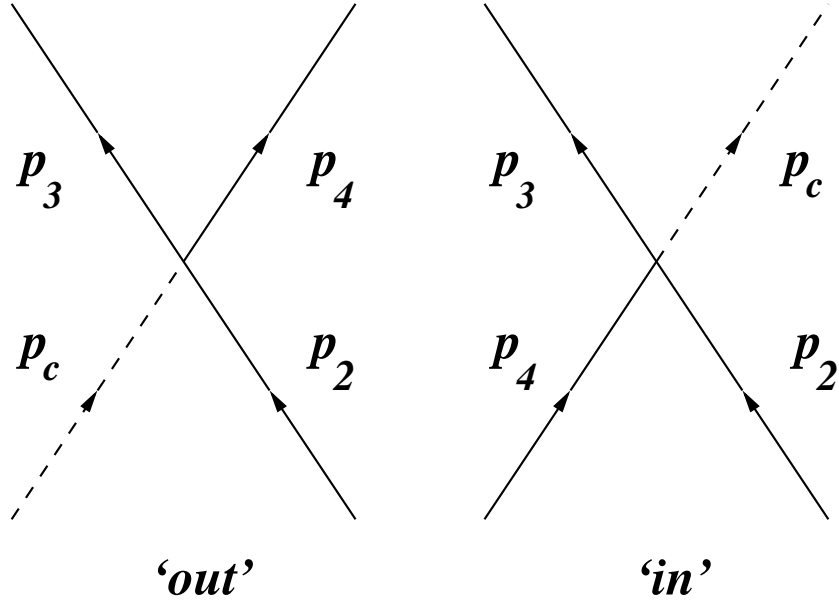, scale=0.9, angle=0}
\vspace{0.8cm}
\caption{\label{C_12}
Schematic illustration of the $C_{12}$ `in' and `out' collision
processes. The dashed line represents a condensate atom and the solid
lines represent thermal atoms.
 }
\end{figure}

\begin{figure}
\centering
\psfig{file=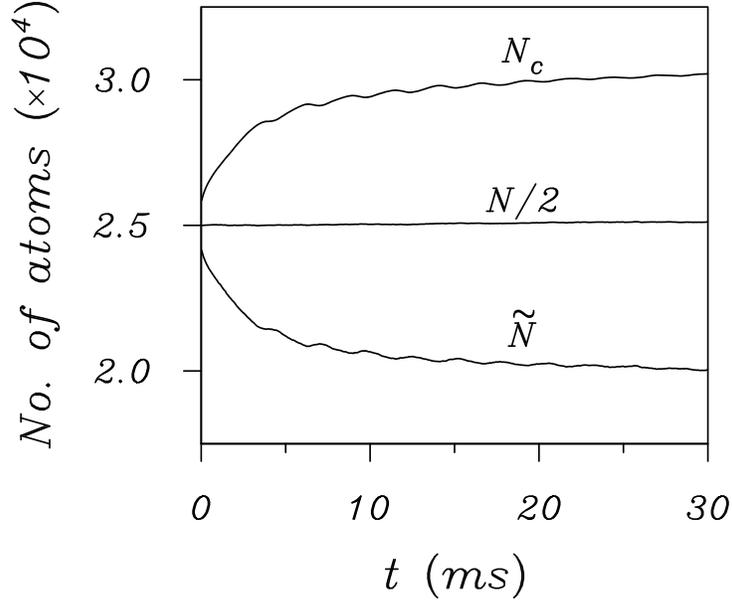, scale=0.52, angle=0, bbllx=10, bblly=145, 
 bburx=540, bbury=605}
\caption{\label{number}
The condensate ($N_c$), noncondensate ($\tilde N$) and total ($N$)
number of atoms as a function of time following a momentum quench. 
The trap has a frequency of
$\nu_0 = 187$ Hz, and contains $5\times 10^4$ $^{87}$Rb atoms at an
initial temperature of $T_0 = 200$ nK.
 }
\end{figure}

\begin{figure}
\centering
\psfig{file=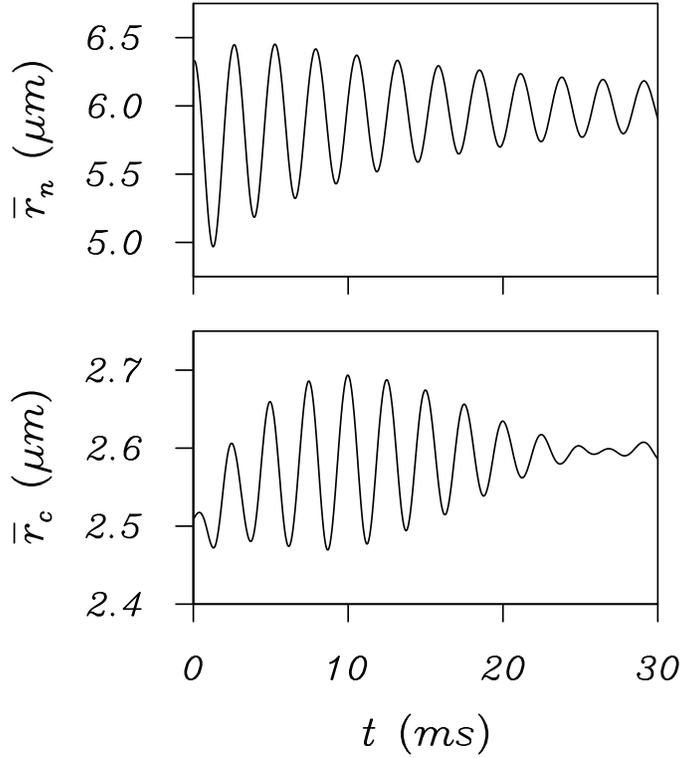, scale=0.52, angle=0, bbllx=45, bblly=65, 
 bburx=540, bbury=640}
\caption{\label{rms_radius}
The rms radius, $\bar r = \sqrt{\langle r^2 \rangle}$, as a function of
time following a momentum quench; $\bar r_n$ is for the thermal cloud 
and $\bar r_c$ for the condensate.
 }
\end{figure}

\begin{figure}
\centering
\psfig{file=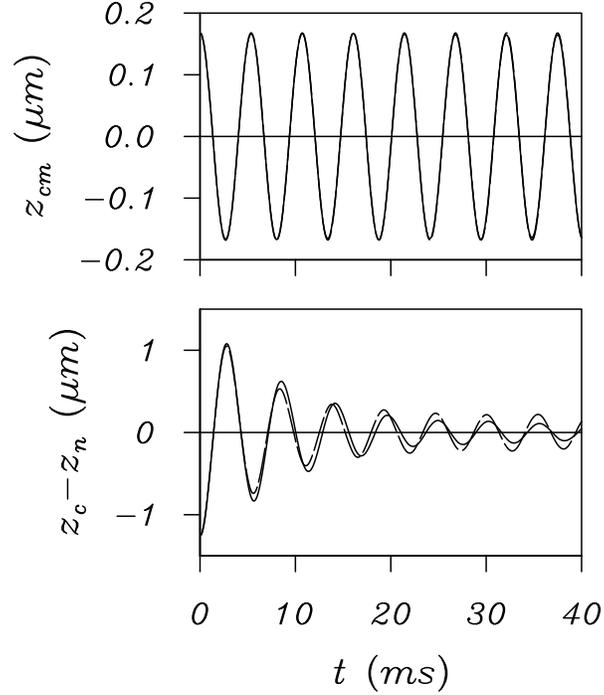, scale=0.47, angle=0, bbllx=25, bblly=110, 
 bburx=510, bbury=690}
\caption{\label{dipole}
The center of mass position, $z_{cm}$, and the displacement between the
mean condensate and noncondensate positions, $z_c - z_n$, as a function
of time. The initial conditions are given in the text. The dashed curves
are in the absence of collisions, while the solid curves include both
$C_{12}$ and $C_{22}$ collisions.
 }
\end{figure}

\begin{figure}
\centering
\psfig{file=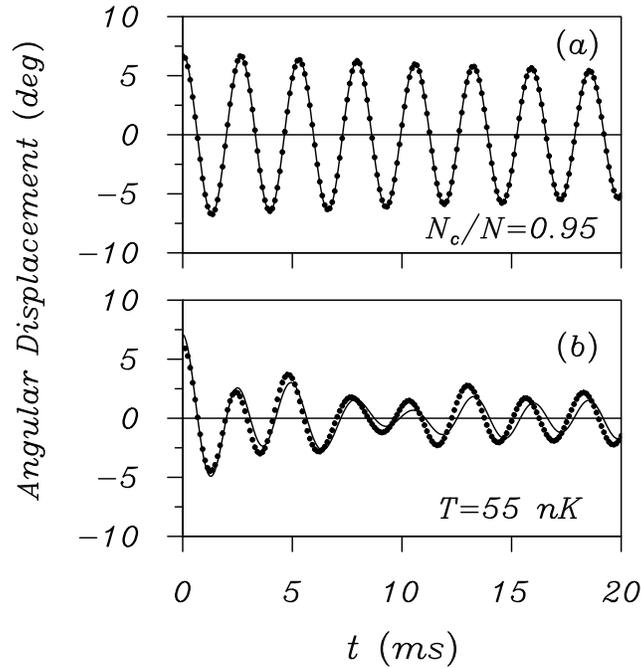, scale=0.45, angle=0, bbllx=40, bblly=110, 
 bburx=570, bbury=690}
\caption{\label{angle55}
Angular displacements of the condensate (a) and thermal cloud (b) as a
function of time. The points are the results of the simulation and the
solid lines are three-mode fits to the data. The results are for the
trap studied experimentally in Ref.\ \protect\cite{marago01}.
}
\end{figure}

\begin{figure}
\centering
\psfig{file=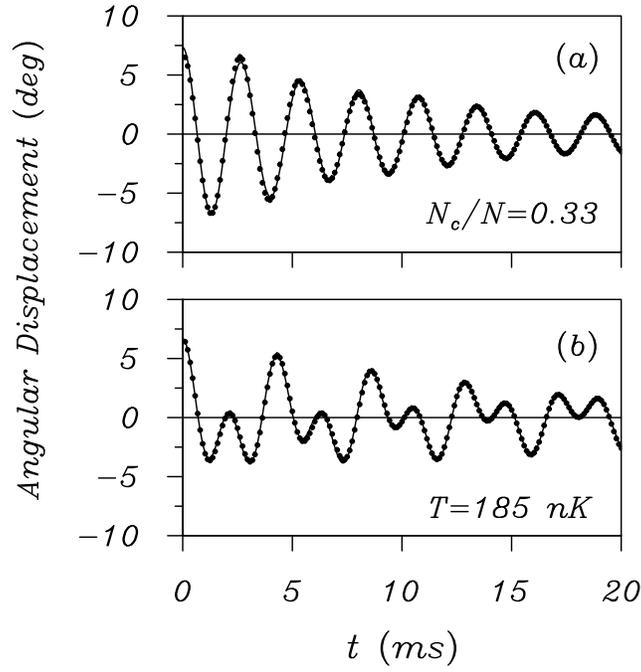, scale=0.45, angle=0, bbllx=40, bblly=110, 
 bburx=570, bbury=690}
\caption{\label{angle185}
As in Fig.~\ref{angle55} but for $T=185$ nK.
 }
\end{figure}

\begin{figure}
\centering
\psfig{file=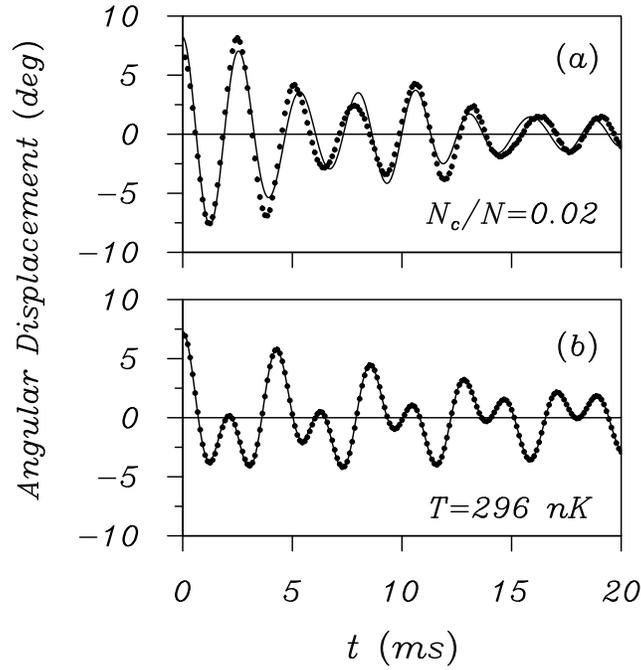, scale=0.45, angle=0, bbllx=40, bblly=110, 
 bburx=570, bbury=690}
\caption{\label{angle296}
As in Fig.~\ref{angle55} but for $T=296$ nK.
 }
\end{figure}

\begin{figure}
\centering
\psfig{file=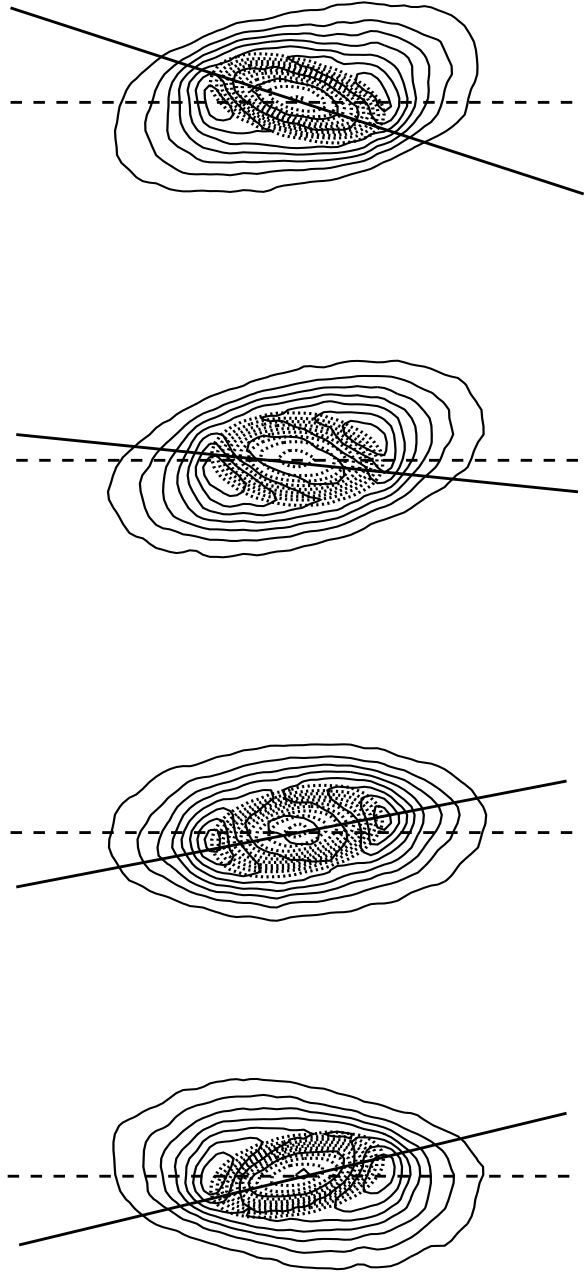, scale=1.2, angle=0, bbllx=25, bblly=30, 
 bburx=225, bbury=465}
\caption{\label{contours}
Density contour plots of the condensate (dotted lines) and thermal cloud
(solid lines) for a sequence of equally spaced time intervals spanning
approximately one-half period of the scissors mode oscillation. The 
earliest time is at the top of the figure. The dashed horizontal line
denotes the symmetry axis of the trapping potential and the straight
solid line indicates the major axis of the condensate at each time step.
 }
\end{figure}

\begin{figure}
\centering
\psfig{file=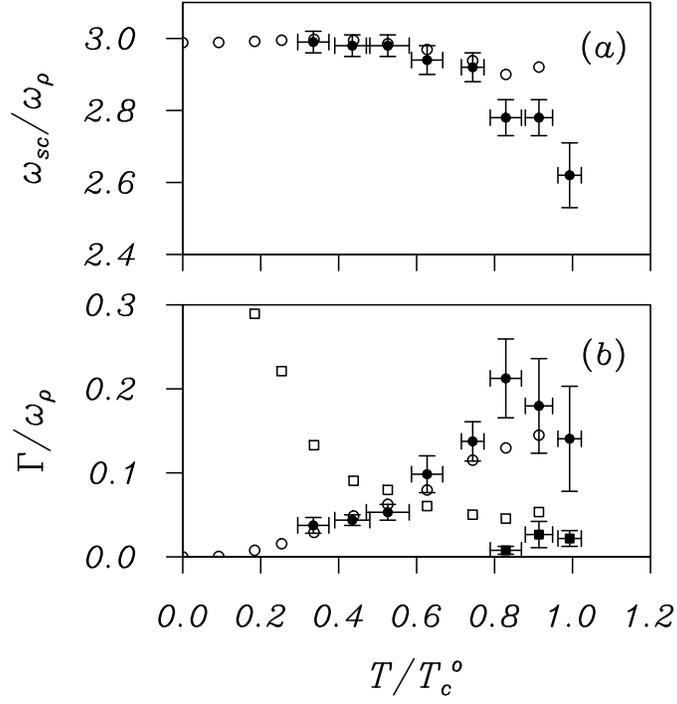, scale=0.48, angle=0, bbllx=50, bblly=105, 
 bburx=570, bbury=665}
\caption{\label{experiment}
 Frequency (a) and damping rate (b) of the scissors modes for a variable
 total number of atoms, intended to simulate the experiments in
 Ref.\ \protect\cite{marago01}. The condensate mode is indicated by open 
 (theory) and solid (experiment) circles. The open squares in (b) show the 
 calculated average 
 damping rate of the two thermal cloud modes, while the solid squares 
 are the corresponding experimental values.
 }
\end{figure}

\begin{figure}
\centering
\psfig{file=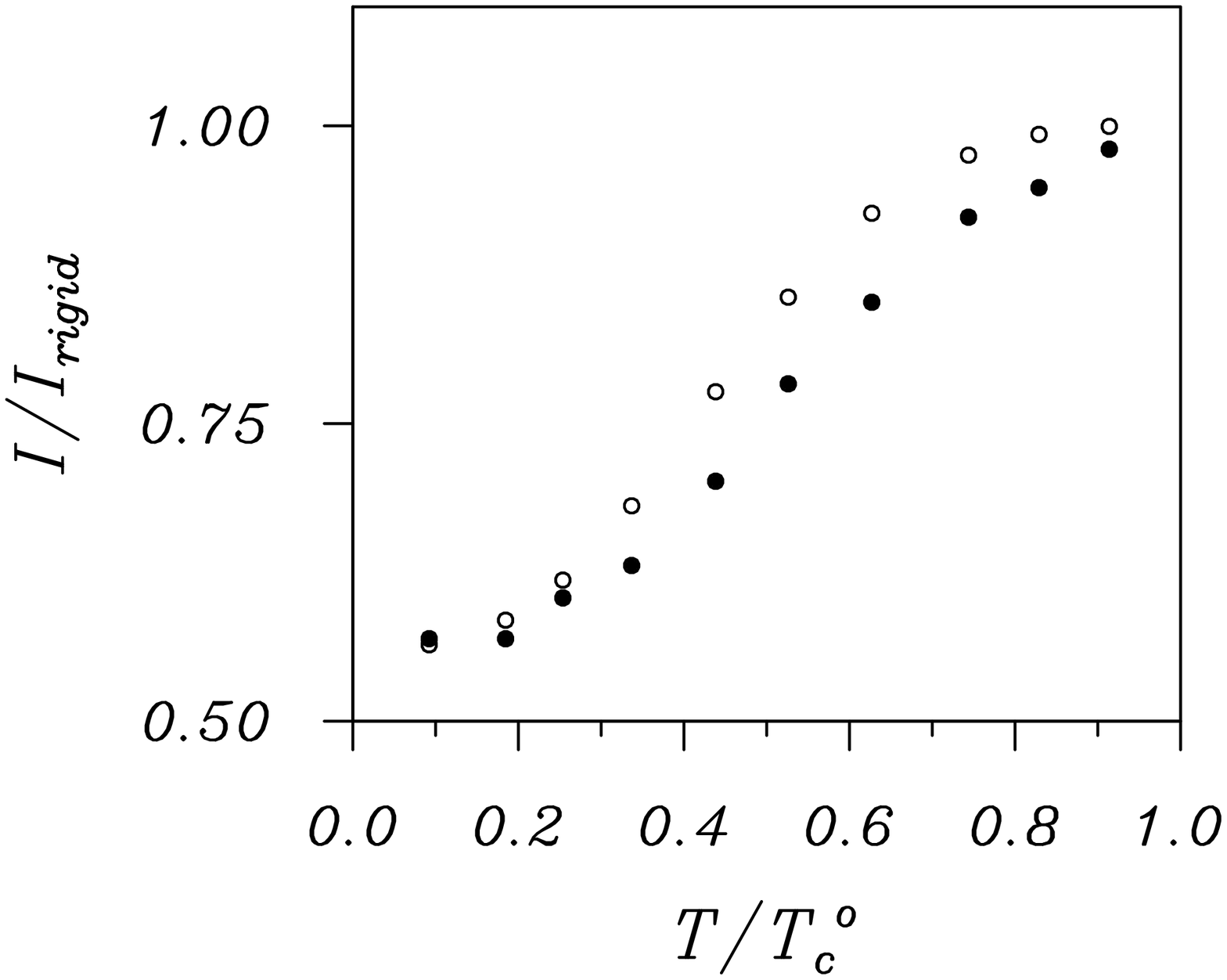, scale=0.5, angle=0, bbllx=25, bblly=140, 
 bburx=575, bbury=610}
\caption{\label{inertia}
The ratio of the moment of inertia of the trapped gas to the rigid body
moment of inertia as a function of temperature. The solid circles are
the result of Eq.~(\ref{eqn30}) while the open circles are obtained
using Eq.~(\ref{eqn33}).
 }
\end{figure}

\begin{references}
\bibitem{footnote1} We note that $\tau_{12}^0$ is defined
 differently in Ref.~\cite{nikuni99}. There, the $\Gamma_{12}^{out}$
 collision rate is divided by $n_{c0}$, rather than by $\tilde n_0$ as
 done here. The latter is a more useful definition for the purpose of
 comparing $C_{12}$ and $C_{22}$ collision rates.
\end{references}
\end{document}